\documentclass[english,nofootinbib,aps,reprint,superscriptaddress,twocolumn]{revtex4-2}
\usepackage[T1]{fontenc}
\usepackage[utf8]{inputenc}
\usepackage{xcolor}
\usepackage{array}
\usepackage{booktabs}
\usepackage{amsmath}
\usepackage{amssymb}
\usepackage{graphicx}
\usepackage{esint}
\PassOptionsToPackage{normalem}{ulem}
\usepackage{ulem}
\usepackage{nicefrac}
\DeclareUnicodeCharacter{03C0}{$\pi$}
\usepackage{amsmath}
\usepackage{cancel}

\makeatletter

\providecolor{lyxadded}{rgb}{0,0,1}
\providecolor{lyxdeleted}{rgb}{1,0,0}

\DeclareRobustCommand{\lyxsout}[1]{\ifx\\#1\else\sout{#1}\fi}

\usepackage[english]{babel}
\usepackage{graphicx}
\usepackage{dcolumn}
\usepackage{bm}
\usepackage{xcolor}

\def\url#1{}
\addto\captionsenglish{%
}

\usepackage{babel}

\providecommand*{\diff}%
{\@ifnextchar^{\DIfF}{\DIfF^{}}}
\def\DIfF^#1{%
	\mathop{\mathrm{\mathstrut d}}%
	\nolimits^{#1}\gobblespace}
\def\gobblespace{%
	\futurelet\diffarg\opspace}
\def\opspace{%
	\let\DiffSpace\!%
	\ifx\diffarg(%
	\let\DiffSpace\relax
	\else
	\ifx\diffarg[%
	\let\DiffSpace\relax
	\else
	\ifx\diffarg\{%
	\let\DiffSpace\relax
	\fi\fi\fi\DiffSpace}

\makeatother

\begin{document}

\title{Wilson polygons and the topology of zero-dimensional systems}
\author{Gen Yin}
\thanks{gen.yin@georgetown.edu}
\affiliation{Department of Physics, Georgetown University, Washington, D.C. 20057,
USA}
\author{Rameswar Bhattacharjee}
\affiliation{Department of Chemistry, Georgetown University, Washington, D.C. 20057, USA}
\author{Thomas Wang}
\affiliation{Department of Physics, Georgetown University, Washington, D.C. 20057,
USA}
\affiliation{Department of Chemistry, Johns Hopkins University, Baltimore, Maryland 21218,
USA}
\author{Miklos Kertesz}
\thanks{kertesz@georgetown.edu}
\affiliation{Department of Chemistry, Georgetown University, Washington, D.C. 20057, USA}

\begin{abstract}
We show that zero-dimensional (0-D) systems can host non-trivial topology analogous to macroscopic topological materials in higher dimensions. Unlike macroscopic periodic systems with translational symmetry, zero-dimensional materials such as molecules, clusters and quantum dots can exhibit discrete rotation symmetry. The eigenstates can thus be grouped into discrete bands and Bloch-like wave functions. Since the symmetry is discrete, the Berry phase and the topological indices must be defined by discrete Wilson polygons. Here, we demonstrate non-trivial $\mathcal{Z}_2$ orders in two representative 0-D molecules, [m]-Cycloparaphenylene and [m]-iso-thianaphthene, where topological transitions occur when modifying the coupling between the repeating units. Similar to macroscopic topological systems in higher dimensions, localized boundary states emerge in composite nanohoops formed by segments that are topologically distinct. This opens up the possibility of non-trivial topological phases in 0-D systems.  
\end{abstract}

\maketitle
Topological band theory was one of the important centerpieces of condensed matter physics for the past several decades. 
It led to the discovery of many topological materials hosting properties that are compelling both for fundamental and application reasons. 
These discoveries include Chern insulators\cite{haldane_model_1988}, topological insulators\cite{fu_topological_2007,roy_topological_2009,moore_topological_2007}, topological superconductors\cite{kitaev_periodic_2009,kitaev_unpaired_2001,qi_time-reversal-invariant_2009,qi_topological_2011} and Weyl and Dirac semimetals\cite{xu_discovery_2015,armitage_weyl_2018}. 
At interfaces between materials with different topological indices, lower-dimensional boundary states emerge, whose existence and robustness are protected by topology.
These boundary states have been recognized as the origin of many important quantized phenomena such as quantum Hall effect\cite{thouless_quantized_1982}, quantum anomalous Hall effect\cite{haldane_model_1988,yu_quantized_2010,chang_experimental_2013,kou_metal--insulator_2015,fox_part-per-million_2018} and quantum spin Hall effect\cite{bernevig_quantum_2006,murakami_spin-hall_2004,kane_z2_2005,konig_quantum_2007}. 
Similar concepts have been applied for other quasi-particle excitations in different periodic systems, resulting in topological phonons\cite{kane_topological_2014}, photons\cite{wang_observation_2009} and magnons\cite{shindou_topological_2013}. 
Recently, organic one-dimensional (1-D) topological systems\cite{pendas_chemical_2019} attracted much attention due to the vast structural and chemical degrees of freedom provided by the well-developed synthesis means. 
Particularly, a $\mathcal{Z}_2$ topological transition has been experimentally identified in a series of ethynylene-linked oligoacene $\pi$-conjugated polymers absorbed on gold surface when varying the acene size\cite{cirera_tailoring_2020,gonzalez-herrero_atomic_2021}. 
Similar topological transitions have been demonstrated for various $\pi$-conjugated polymers as a function of external strain\cite{bhattacharjee_topological_2024}. 
The aforementioned topological classification is mainly based on the band theory of periodic materials owing to their translational symmetry. 
These systems are usually both macroscopic and periodic, and the crystal momentum, $\hbar\mathbf{k}$, is therefore continuous in the reciprocal space. 
The Berry (or Zak) phase, the Berry connection and the Berry curvature are thus defined as continuous functions of $\mathbf{k}$\cite{zak_berrys_1989,berry_quantal_1984}. 
Specifically, the Berry connection can be taken as a gauge field, whose integration along a closed Wilson loop provides a gauge-invariant Berry phase\cite{wilson_confinement_1974}. 
In some cases, the continuous Wilson loop can be discretized, which is important in numerical calculations of polarization\cite{king-smith_theory_1993}, the Fermi-surface Hall conductivity\cite{wang_fermi-surface_2007,yu_discrete_2021}, optical resonses\cite{wang_generalized_2022} and topological indices\cite{fukui_chern_2005}. 
These loops formed by discrete Wilson links are also known as Wilson polygons\cite{cherednikov_loop_2012}. 
Although well established, the Wilson polygon approach is often considered an approximation to handle the continuous spectra of macroscopic topological materials in 1D, 2D or 3D. 
Here, we show that Wilson polygons define the topological indices of zero-dimensional (0-D) systems\cite{bhattacharjee_quinonoid_2025}. 
This definition is not an approximation of a continuous limit, and is naturally robust in numerical calculations. 
\begin{figure*}
\begin{centering}
\includegraphics[width=0.9\textwidth]{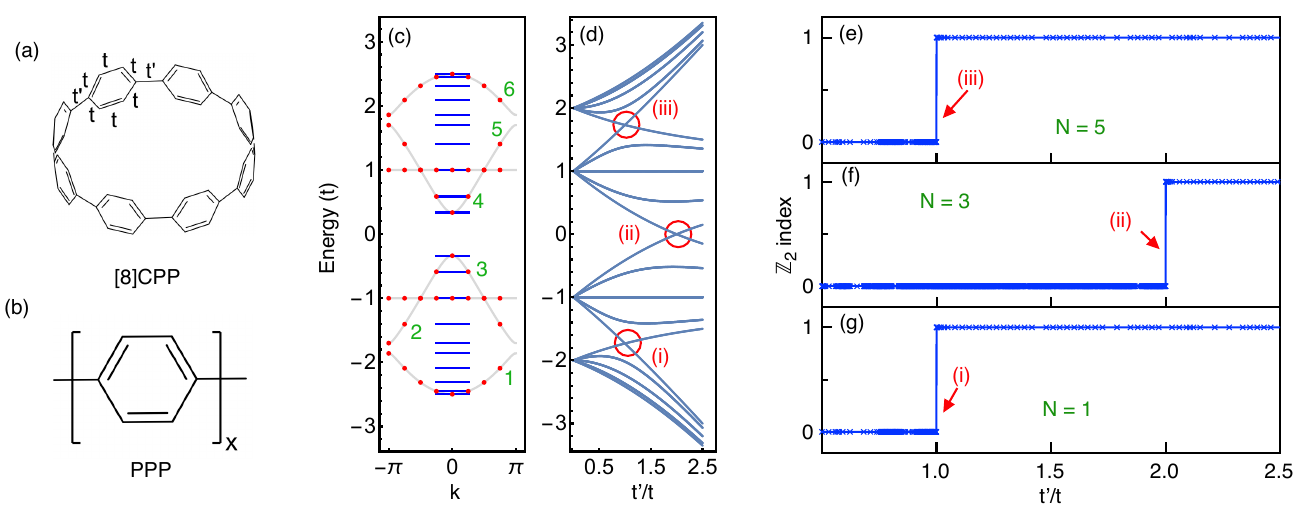}
\par\end{centering}
\caption{\textbf{The $\mathcal{Z}_2$ transitions in $\textrm{[m]-CPP}$.}  (a) The atomic structure of a $\textrm{[8]-CPP}$. Here, $t$ is the C-C hopping amplitude within each benzene ring, whereas $t'$ labels the hopping amplitude between neighboring units. (b) The unit cell of a PPP polymer corresponding to the case of $\textrm{[m]-CPP}$ when $m\rightarrow\infty$. (c) The mapping of MOs in [8]-CPP onto the continuous bands of a PPP at $t'=t$. The short horizontal blue lines denote the discrete energy levels of the MOs, and their corresponding mapping positions in the continuous bands are denoted by the red solid points. The gray solid lines illustrate the continuous bands of the 1-D polymer PPP. (d) The evolution of the energy levels of $\textrm{[8]-CPP}$ when modulating $t'$. Three crossings are captured. (e-g) The quantized transition of the $\mathcal{Z}_{2,N}$ index as a function of $t'$, corresponding to the three crossings shown in (d). \label{fig:CPP}}
\end{figure*}
Specifically, we demonstrate the topological order in two 0-D molecules: the [m]cycloparaphenylene\cite{darzi_dynamic_2015} ($\textrm{[m]-CPP}$) and [m]-iso-thianaphthene ($\textrm{[m]-ITN}$). 
Since the molecules are circular, they are also known as `nanohoops'.
We use modified H\"uckel models including only the first-neighbor hopping integrals for both molecules. 
Due to the time-reversal symmetry and the small spin-orbit coupling in both systems, spin degeneracy is assumed for all orbitals. 
This is similar to the Su-Schrieffer-Heeger model of 1-D polyacetylene\cite{su_solitons_1979}. 
We first demonstrate the concept using a simplified model of $\textrm{[m]-CPP}$, as shown in Fig. \ref{fig:CPP}(a). 
This molecule consists of $m$ para-linked phenyl groups, similar to poly-para-phenylene, PPP, the polymeric semiconducting analogue, as shown in Fig. \ref{fig:CPP}(b). 
Here, $t$ describes the carbon-carbon (CC) hopping within each repeating unit, whereas $t'$ represents the coupling between the adjacent ones [Fig.\ref{fig:CPP}(a)]. 
Since the molecule is circular, it has $\mathcal{C}_m$ rotational symmetry, and the molecular orbitals (MOs) are given by the standing-wave solutions propagating along the nanohoop.
The local wavefunction associated to one repeating unit is different from those associated to other ones only by an overall phase. 
This phase is given by the eigenvalues of discrete rotation: $e^{ikm}=1$, where $k=\frac{2\pi l}{m}$, and $l=0,1,2,\cdots,m-1$. 
At the limit of $m\rightarrow\infty$, the discrete $k$ points become continuous and the rotational symmetry goes back to the translational symmetry in a 1-D lattice. 
At this limit, the Bloch theorem is restored, and the 0-D  $\textrm{[m]-CPP}$ molecule becomes the macroscopic 1-D poly-para-phenylene (PPP). 
We can therefore precisely map the MOs to the PPP bands, as shown in Fig. \ref{fig:CPP}(c). 
The single parameter $\nicefrac{t'}{t}$ in the H\"uckel model significantly modulates the MO energies as shown in Fig. \ref{fig:CPP}(d).
For a realistic model of PPP or $\textrm{[m]-CPP}$, $\nicefrac{t'}{t}<1$ is restricted, because the carbon-carbon bonds connecting benzene rings are longer, corresponding to an overall structure that is strictly `aromatic', meaning that the $\pi$-electrons are more localized on the benzene rings and less on the interring bonds\cite{kertesz_conjugated_2005}.
Here we modulate $\nicefrac{t'}{t}$ in an unrealistic range to demonstrate the concept, and more realistic systems will be introduced later. 
When modulating $0\le\nicefrac{t'}{t}\le2.5$, the model captures three crossing of MOs, which are labeled by (i)-(iii) in Fig. \ref{fig:CPP}(d). 
Crossings (i) and (iii) are at $\nicefrac{t'}{t}=1$, which can be mapped to the two band crossings at $k=-\pi$, as shown in Fig. \ref{fig:CPP}(c). 
Crossing (ii) is captured at $\nicefrac{t'}{t}=2$, which is mapped to the gap between Bands 3 and 4, and is always occurring at $k=0$ [Fig. \ref{fig:CPP}(c)]. 
Considering the fact that each carbon atom contributes one electron to the $\pi$ system, Crossing (ii) corresponds to the gap between highest occupied MO (HOMO) and the lowest unoccupied MO (LUMO). 

These crossings of energy levels indeed correspond to $\mathcal{Z}_2$ topological transitions.
Due to the finite size of the nanohoop, the manifold is naturally discrete, and the Berry phase can only be defined on a Wilson polygon: $\gamma_n = -\mathrm{Im}\, \ln\left( \prod_{l=1}^{m} \langle k_l, n \mid k_{l+1}, n \rangle \right)$, with $k_{m+1} = k_1$, where $|k_l,n\rangle$ denotes the periodic part of the $l$-th discrete MO sampled along the $n$-th band. 
Summation of the Berry phase up to the $N$-th band provides the corresponding index: $\mathcal{Z}_{2,N} = \sum_{n=1}^{N} \frac{\gamma_n}{\pi} \quad\text{(mod 2)}$.
This is equivalent to  
\begin{equation}
\mathcal{Z}_{2,N} = -\frac{1}{\pi} \, \mathrm{Im}\, \ln\left( \prod_{n=1}^{N} \prod_{(l\in\textrm{band }n)}^m \langle k_l, n \mid k_{l+1}, n \rangle \right),
\label{eq:z2_index}
\end{equation}
where index $l$ goes through the $n$-th band, and the natural logarithm is defined on the branch $(-\pi,\pi]$. 
The $\mathcal{Z}_2$ index as a function of $\nicefrac{t'}{t}$ is shown in Figs. \ref{fig:CPP}(e-g). 
Here, mod 2 is always taken after evaluating Eq. \ref{eq:z2_index}.
Sharp steps are captured numerically at $\nicefrac{t'}{t}=1$ and $\nicefrac{t'}{t}=2$, respectively, exactly coinciding with the level crossings of the MOs shown in Fig. \ref{fig:CPP}(d). 
The discussion here describes only the ground-state physics, and thus only considers the Abelian Berry connection. 
The sharp transition steps captured in Figs. \ref{fig:CPP}(e-g) suggest good numerical stability near degeneracy.
\begin{figure*}
\begin{centering}
\includegraphics[width=1\textwidth]{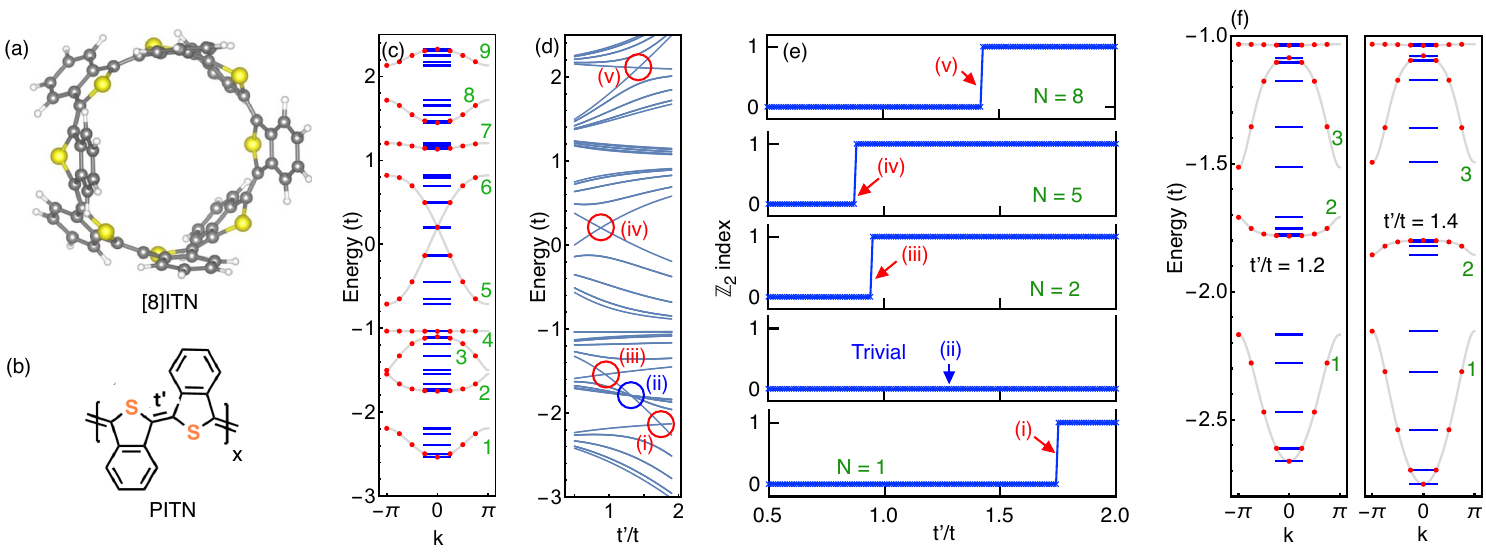}
\par\end{centering}
\caption{\textbf{The $\mathcal{Z}_2$ transitions in $\textrm{[m]-ITN}$.}  (a) An illustration of the structure of $\textrm{[8]-ITN}$, where the repeating units take a alternative structure after structure optimization. (b) The unit cell of a PITN polymer corresponding to the case of [m]ITN when $m\rightarrow\infty$. (c) The mapping of MOs in $\textrm{[8]-ITN}$ onto the continuous bands of a PITN similar to that shown in Fig.\ref{fig:CPP}.  Here $\frac{t'}{t}=0.88$, corresponding to the crossing of the HOMO and LUMO at $\Gamma$. (d) The evolution of the energy levels of $\textrm{[8]-ITN}$ when modulating $t'$. Five crossings (i-v) are captured. (e) The $\mathcal{Z}_{2,N}$ indices as a function of $t'$, corresponding to the five crossings shown in (d). The transition in (ii) is trivial, whereas all others are topologically non-trivial. (f) The explanation of the trivial transition for Case (ii). The left and the right panels show the different band curvature before (left) and after (right) the crossing. \label{fig:ITN}}
\end{figure*}
\begin{figure}
\begin{centering}
\includegraphics[width=1\columnwidth]{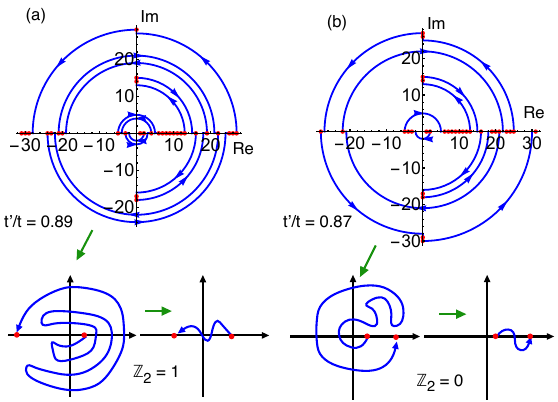}
\par\end{centering}
\caption{\textbf{The geometric meaning of the discrete Berry phase and the $\mathcal{Z}_2$ index.}  The solid red points illustrate the accumulative Berry phase when calculating the index $\mathcal{Z}_{2,N=5}$ of [6]-ITN. The magnitude of each point is normalized to the MO index when looping through the $6\times5=30$ MOs within the first five bands. The blue arcs with arrows indicate the order of Berry phase accumulation. (a) The topologically non-trivial case of $\mathcal{Z}_{2,N=5}=1$ corresponds to a net Berry phase of $\pi$. Smooth gauge transformations can deform the path into a line segment that must contain the origin. (b) The trivial case where the total Berry phase is $0$, such that the path can deform into a line segment excluding the origin. \label{fig:ZakLoops}}
\end{figure}

The non-trivial $\mathcal{Z}_2$ orders in 0-D $\textrm{[m]-CPP}$ are protected by the inversion symmetry with the geometric center of the nanohoop being the inversion center. 
This is guaranteed when $m$ is even, where one can map the azimuthal angle $\phi\rightarrow\phi+\pi$ in a cylindrical coordinate. 
The $\mathcal{Z}_2$ index can therefore also be dictated by the inversion eigenvalues at $k=0$ and $k=-\pi$, which is similar to a 1-D topological crystalline insulator\cite{fu_topological_2011,slager_space_2013}.  
When $m$ is odd, however, the gaps at $k=-\pi$ cannot be sampled by the discrete MOs. 
In real space, the geometric center of the nanohoop in this case is no longer an inversion center. 
The calculated $\mathcal{Z}_2$ index is thus an arbitrary value. 
Nevertheless, the Berry phase along the Wilson polygon remains gauge invariant. 
At the limit of $m\rightarrow\infty$, the even-odd subtlety vanishes as expected. 
Details of this discussion can be found in Supplemental Material Sec. I\cite{yin_supplemental_2025}.
Although the demonstration here is using a tight-binding model, the formalism can easily be generalized for a full-band first-principles model after a proper basis rotation restores the cyclic periodicity of the Hamiltonian. 
More details can be found in the Supplemental Material Sec. II \cite{yin_supplemental_2025}, where we apply Eq. \ref{eq:z2_index} on the Fock and overlap matrices obtained by the quantum chemistry package ORCA\cite{neese_software_2025, becke_densityfunctional_1993,lee_development_1988,hehre_selfconsistent_1972}.
Although topological transitions must coincide with the crossing of some energy levels, simply having the crossing of MOs does not necessarily suggest topological transition. 
This can be seen in the spectrum of $\textrm{[m]-ITN}$, which consists of $m$ repeating iso-thia-naphthene units [Fig. \ref{fig:ITN}(a)]. 
At the limit of $m\rightarrow\infty$, the nanohoop becomes PITN, a polymer known to host a small band gap close to $1\thinspace\textrm{eV}$\cite{bredas_towards_1986}, as illustrated in Fig. \ref{fig:ITN}(b).
The corresponding tight-binding model has two more parameters: the sulfur on-site energy and the S-C coupling. 
Both were chosen according to traditional parametrization as $\epsilon_\textrm{S}=1.11t$ and $t_{\textrm{C-S}}=0.69t$, respectively\cite{carissan_huckel-lewis_2008}. 
Each sulfur atom contributes two $\pi$-electrons in this hybridization. 
The coupling between the units remains $t'$, and again $\nicefrac{t'}{t}$ is the single parameter of the presented model. 
As shown in Fig. \ref{fig:ITN}(c), the discrete spectrum of $\textrm{$\textrm{[8]-ITN}$}$ can be exactly mapped to the continuous spectrum of PITN. 
These quantized MOs also experience multiple crossings when the hopping element $t'$ is modulated, which are labeled by (i)-(v) in Fig. \ref{fig:ITN}(d). 
Among these five crossings, only (i), (iii), (iv) and (v) correspond to a $\mathcal{Z}_2$ topological transition 
as shown in Fig. \ref{fig:ITN}(e). 
These four transitions can be mapped to the PITN band gaps between pairs of (1, 2), (2, 3), (5, 6) and (8, 9), respectively. 
The crossing at (ii) is unique and does not correspond to any topological transition. 
This crossing is mapped to the 2nd band of PITN, which changes from convex to concave when increasing $\nicefrac{t'}{t}$ as shown in Fig. \ref{fig:ITN}(f). 
Although the MOs are indeed crossing, no band gap is closing at the continuous limit. 
This gives a trivial crossing of the MOs, which is consistent with the corresponding behavior of the $\mathcal{Z}_2$ index shown in Fig. \ref{fig:ITN}(e) for Case (ii). 
Nevertheless, this trivial crossing occurs deep in the occupied MOs since we have ten $\pi$ electrons in the system. 
The HOMO-LUMO crossing is always topologically non-trivial as shown in Case (iv). 
The topological indices of 0-D nanohoops defined on Wilson polygons are numerically robust.
This can be seen by examining the accumulative Berry phase when looping through the edges of the Wilson polygon. 
In Fig. \ref{fig:ZakLoops} we illustrate the accumulative total phase of the complex product within the parenthesis in Eq. \ref{eq:z2_index}. 
For the case of $\mathcal{Z}_{2,N=5}$ and $\textrm{[6]-ITN}$, this involves $5\times6=30$ steps. 
The polar angle demonstrates the corresponding instantaneous phase, whereas the magnitude indicates the indices of the steps. 
In fact, the specific path of the instantaneous Berry phase is not unique, and is determined by the choice of gauge, which is usually arbitrarily assigned by the numerical solver. 
As shown in the upper panels of Figs. \ref{fig:ZakLoops}(a-b), the specific path of phase accumulation is usually abrupt and not differentiable. 
Nevertheless, one can indeed choose a continuous gauge and smoothly deform the paths into continuous curves that are topologically equivalent. 
This transformation can simplify the path of phase accumulation as if we are tightening a string. 
When $\mathcal{Z}_2=0$, the beginning and the ending Berry phase remain equal (in our case, zero), such that the path can be transformed to a line excluding the origin, as illustrated in the lower panels of Fig. \ref{fig:ZakLoops}(b). 
On the other hand, when $\mathcal{Z}_2=1$, the ending Berry phase is $-\pi$, whereas the beginning phase is $0$. 
The shortest path connecting these two points therefore must contain the origin, as illustrated in Fig. \ref{fig:ZakLoops}(a). 
%
%

\begin{figure}
\begin{centering}
\includegraphics[width=1\columnwidth]{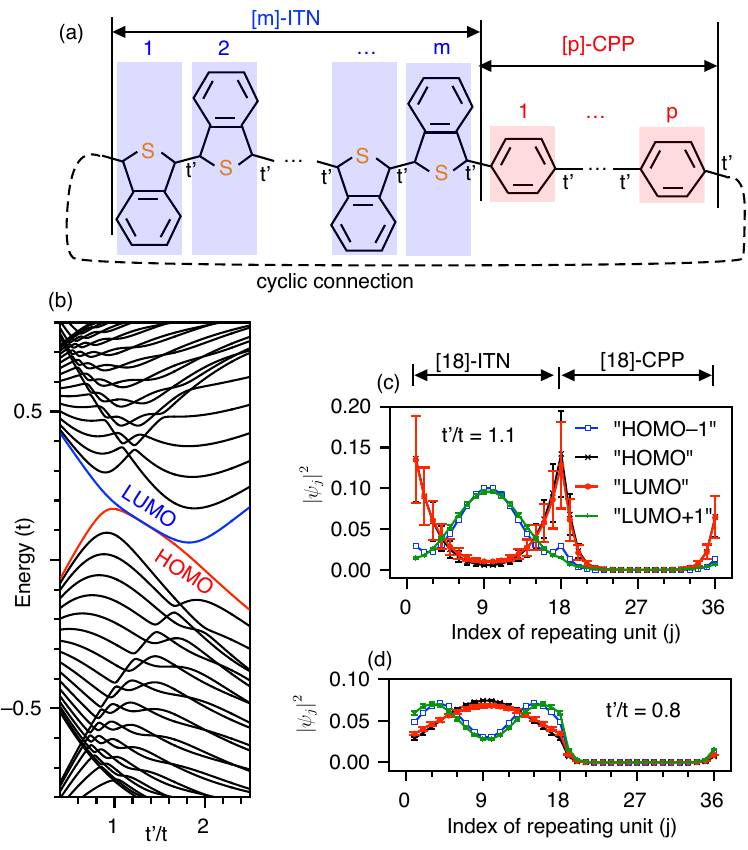}
\par\end{centering}
\caption{\textbf{The topological boundary state in 0-D composite topological nanohoops.} (a) The structure of cyclic $\textrm{[m]-ITN-[p]-CPP}$. The dashed line denotes the C-C bond connecting the first unit of the [m]-ITN segment and the last one in the [p]-CPP segment. (b) The spectrum near the HOMO-LUMO gap of $\textrm{[18]-ITN-[18]-CPP}$ when modifying the single parameter $\nicefrac{t'}{t}$. (c) The average electron density resolved on the repeating units when the two segments are topologically distinct. The error bars denote the standard deviation among $200$ on-site white noise samples. (d) The electron density similar to (c) but when the two segments are topologically equivalent. \label{fig:compositNanohoops}}
\end{figure}
The $\mathcal{Z}_2$ index of a nanohoop can be taken as a hallmark of in-gap boundary states when two topologically distinct $\pi$-bond systems are forming a composite 0-D nanohoop. 
This is similar to the bulk-boundary correspondence for topological materials in higher dimensions.
The boundary state can be demonstrated in $\textrm{[m]-ITN-[p]-CPP}$ whose structure is illustrated in Fig. \ref{fig:compositNanohoops}(a), where the full cyclic nanohoop contains $m$ repeating units in the [m]-ITN segment and $p$ ones in the [p]-CPP segment\cite{bhattacharjee_quinonoid_2025}. 
Such composite nanohoops can also be modeled by the aforementioned H\"uckel model with a single parameter $t'$ denoting the hopping between adjacent repeating units. 
The energy of the MOs at different values of $\nicefrac{t'}{t}$ are illustrated by the solid lines in Fig. \ref{fig:compositNanohoops}(b). 
Due to the $\mathcal{Z}_{2,N=5}$ transition shown in Fig. \ref{fig:ITN}(e), the $\textrm{$\textrm{[8]-ITN}$}$ molecule is topologically nontrivial when $\nicefrac{t'}{t}>\sim0.88$. 
On the other hand, the $\textrm{$\textrm{[8]-CPP}$}$ molecule is topologically trivial when $\nicefrac{t'}{t}<2$ as shown in Fig. \ref{fig:CPP}(f). 
In fact, these transition conditions remain unchanged for arbitrary $m$ and $p$ values, since the HOMO-LUMO gap at $k=0$ is always sampled. 
For example, the two segments of $\textrm{[18]-ITN-[18]-CPP}$ must be topologically distinct if $0.88\leqslant\nicefrac{t'}{t}\leqslant2$.
This is manifested as the merging of the HOMO and LUMO, as illustrated in Fig.\ref{fig:compositNanohoops}(b). 
When this occurs, the electron density along the repeating units are sharply peaked at the interfaces, as shown in Fig. \ref{fig:compositNanohoops}(c). 
Although sharply peaked, the two interfaces can couple due to the small size of the nanohoop, opening a finite gap in the spectrum. 
This explains the splitting of HOMO and LUMO near $\nicefrac{t'}{t}\approx0.88$ and $\approx2.0$ captured in Fig. \ref{fig:compositNanohoops}(b). 
The size of this gap is determined by the penetration depth of the boundary states, which is sensitive to $\frac{t'}{t}$. 
This depth diverges near the critical point ($\frac{t'}{t}\approx0.88$), suggesting a 2nd-order-like phase transition. 
More details can be found in Sec. III of the Supplemental Material\cite{yin_supplemental_2025}. 
When the topological indices of both segments are identical, the HOMO and LUMO resemble the particle-in-a-box ground state as shown in Fig. \ref{fig:compositNanohoops}(d). 
Notably, the LUMO+1 and HOMO-1 states behave normally as excited standing waves in this case.
However, when the system becomes topologically non-trivial, HOMO-1 and LUMO+1 change to the profiles of a typical ground state, while the HOMO and LUMO are sharply peaked at the interfaces as shown in Fig. \ref{fig:compositNanohoops}(c). 
This suggests that the topological transition is indeed a global change for the full spectrum. 
Although the quantization of the $\mathcal{Z}_2$ index discussed here is protected by the inversion symmetry, noise breaking such symmetry does not destroy the boundary states at the interface. 
Here we consider an on-site uncorrelated noise $\delta\epsilon\in[-w,w]$ with $\frac{w}{t}=0.02$ on all C and S sites. 
The standard deviations of the electron density due to $200$ white-noise samples are illustrated by the error bars in Figs. \ref{fig:compositNanohoops}(c) and (d). 
This demonstrates the robustness of the boundary states due to a thermal noise equivalent to $\sim600\thinspace\textrm{K}$. 
More importantly, since chemical bonds near the interface are passivated, the localization of the boundary state can be seen as an experimental hallmark of the topological order. 
This rules out the local density of states induced by dangling bonds in 1-D systems such as topological nanowires. 
We propose optical and transport experiments to detect the topological transitions in these 0-D nanohoops of delocalized $\pi$-electron systems. 
As demonstrated by the three examples shown above, the energy scale corresponding to the topological transitions can be intricately modulated by the specific structure. 
In the case of $\textrm{[8]-CPP}$, the $\mathcal{Z}_2$ transition crossing the HOMO and LUMO occurs at $\nicefrac{t'}{t}=2$, which is difficult to achieve experimentally. 
Distinct from $\textrm{[8]-CPP}$, the $\mathcal{Z}_2$ transition (iv) in $\textrm{$\textrm{[8]-ITN}$}$ occurs at $\nicefrac{t'}{t}\sim0.88$, which is closer to the realistic value. 
Similarly, the merging HOMO-LUMO gap in the composite nanohoop $\textrm{[m]-ITN-[p]-CPP}$ also occurs when $\nicefrac{t'}{t}\sim1$. 
The transition in these two cases are likely to occur when deformation \cite{bhattacharjee_continuous_2024} is induced through mechanical pulling in a break junction\cite{reed_conductance_1997,xu_measurement_2003}.
For example, assuming the orbital decay $t'\approx te^{-\lambda(d'-d)}$ with a decay rate $\lambda=2.5\thinspace\textrm{\AA}^{-1}$, the transition of [p]-ITN should occur when the inter-unit
C-C bond is stretched by $\delta d'=-\frac{ln(0.88)}{2.5}\approx0.05\thinspace\textrm{\AA}$.
Optical spectroscopy or ballistic transport may capture the abrupt change in the MOs. 
In fact, the bond strength between adjacent repeating units in similar organic $\pi$-electron systems can be fine-tuned by changing the long-range repulsion between the adjacent repeating units. 
This is achievable through structure engineering\cite{hermann_conjugated_2021}, corresponding to a vast degree of freedom offered by the frontier of synthetic chemistry. 

\emph{Data Availability Statement:} The data that support the findings of this article are openly available\cite{yin_2025_15313638}. 

\emph{Acknowledgments:} This paper is based upon work supported by the National Science Foundation (US) under Grant \# ECCS-2151809 (GY) and DMR-2440337 (GY). Work by TW was supported by the REU site hosted at Georgetown University supported by NSF(US) under Grant \# DMR-2349397. This work used Bridges-2 at Pittsburgh Supercomputing Center through allocation PHY230018 from the Advanced Cyber infrastructure Coordination Ecosystem Services \& Support (ACCESS) program, which is supported by National Science Foundation (US) grants \#2138259, \#2138286, \#2138307, \#2137603, and \#2138296. This material is partially based on work supported by the U.S. Department of Energy, Office of Science, Office of Basic Energy Sciences under award number DE-SC-0019017 (MK). 

\bibliographystyle{apsRevNoPublisherFullAuthorFullTitle}
\bibliography{references}

\clearpage

\section*{Supplemental Material}

\section{Protection of integer $\mathcal{Z}_2$ index due to inversion symmetry}

The two nanohoops discussed in the main text [m]CPP and [p]ITN are assumed to have ${\cal C}_{m,p}$ discrete rotational symmetry, mapping site $j$ to $j+m$. 
This allows us to associate molecular orbitals with a set of energy bands by Fourier transformation. 
The symmetry protecting the integer values of the $\mathcal{Z}_2$ index is the inversion symmetry mapping $r\rightarrow-r$. 
That means the $j\text{-th}$ repeating unit can be mapped to the $(m-j)\text{-th}$ unit, with $j$ counting the repeating units along the ring. 
In the reciprocal space this maps the eigenstates at $k_l$ to $-k_l$ up to a phase: ${\cal P} |k_l\rangle =\xi_k|-k_l\rangle$, where $\cal P$ is the inversion operator and $\xi_k=e^{i\theta(k_l)}$ is the eigenvalue of $\cal P$. 
Since ${\cal P}^2=1$, we must have $\theta(k)+\theta(-k)=2n\pi$ where $n$ is an integer. 
For arbitrary $0<k<\pi$, the value of $\theta(k)$ is determined by the gauge choice. 
However, the high-symmetry points at $k=0$ and $k=\pi$ are special since $k=-k\bmod 2\pi$, which guarantees $\xi_0^2=\xi_\pi^2=1$. 
Since $\cal P$ is unitary, we have
\begin{equation}
\langle k_l \mid k_{l+1} \rangle = e^{i[\theta(k_{l+1}) - \theta(k_l)]} \langle -k_l \mid -k_{l+1} \rangle.
\end{equation}
When $m$ is even, we can group the complex overlaps along the Wilson polygon as
\begin{widetext}
\begin{align}\mathcal{Z}_{2}= & -\frac{1}{\pi}\,\mathrm{Im}\,\ln\left(\prod_{-\pi\le k<0}\prod_{0\le k<\pi}\langle k\mid k+\delta k\rangle\right) \label{eq:EvenCase} \\
= & -\frac{1}{\pi}\,\mathrm{Im}\,\ln\left(\prod_{-\pi\le k<0}\langle k\mid k+\delta k\rangle\prod_{0\le k<\pi}e^{i[\theta(\pi)\cancel{-\theta(\pi-\delta k)}\cancel{+\theta(\pi-\delta k)}-\cdots\cancel{+\theta(\delta k)}-\theta(0)]}\langle-k\mid-k-\delta k\rangle\right)\\
= & -\frac{1}{\pi}\,\mathrm{Im}\,\ln\left(e^{i[\theta(\pi)-\theta(0)]}\prod_{-\pi\le k<0}\langle k\mid k+\delta k\rangle\prod_{-\pi<k\le0}\langle k\mid k-\delta k\rangle\right)\\
= & -\frac{1}{\pi}\,\mathrm{Im}\,\ln\left(e^{i[\theta(\pi)-\theta(0)]}\underbrace{\prod_{-\pi\le k<0}\langle k\mid k+\delta k\rangle\prod_{-\pi\le k<0}\langle k+\delta k\mid k\rangle}_{\text{real}}\right)\\
= & -\frac{1}{\pi}\thinspace\textrm{Im}\thinspace\ln\frac{\xi_{\pi}}{\xi_{0}}=\begin{cases}
0 & \text{if }\xi_{\pi}\xi_{0}=+1\\
1 & \text{if }\xi_{\pi}\xi_{0}=-1
\end{cases}
\end{align}
\end{widetext}
On the other hand, when $m$ is odd, the separation of Wilson polygon always has an unpaired overlap:
\begin{widetext}
\begin{align}
\mathcal{Z}_{2}= & -\frac{1}{\pi}\,\mathrm{Im}\,\ln\left(\langle\pi-\delta k\mid-\pi+\delta k\rangle\prod_{(-\pi+\delta k)\le k<0}\prod_{0\le k<(\pi-\delta k)}\langle k\mid k+\delta k\rangle\right)  \\
= & -\frac{1}{\pi}\,\mathrm{Im}\,\ln\left(\langle\pi-\delta k\mid-\pi+\delta k\rangle\prod_{(-\pi+\delta k)\le k<0}\langle k\mid k+\delta k\rangle\right. \\ 
 &\left.\prod_{0\le k<(\pi-\delta k)}e^{i[\theta(\pi-\delta k)\cancel{-\theta(\pi-2\delta k)}\cancel{+\theta(\pi-2\delta k)}-\cdots\cancel{+\theta(\delta k)}-\theta(0)]}\langle-k\mid-k-\delta k\rangle\right)\\
= & -\frac{1}{\pi}\,\mathrm{Im}\,\ln\left(\langle\pi-\delta k\mid-\pi+\delta k\rangle e^{i[\theta(\pi-\delta k)-\theta(0)]}\prod_{(-\pi+\delta k)\le k<0}\langle k\mid k+\delta k\rangle\prod_{(-\pi+\delta k)<k\le0}\langle k\mid k-\delta k\rangle\right)\\
= & -\frac{1}{\pi}\,\mathrm{Im}\,\ln\left(\langle\pi-\delta k\mid-\pi+\delta k\rangle e^{i[\theta(\pi-\delta k)-\theta(0)]}\underbrace{\prod_{(-\pi+\delta k)\le k<0}\langle k\mid k+\delta k\rangle\prod_{(-\pi+\delta k)\le k<0}\langle k+\delta k\mid k\rangle}_{\text{real}}\right)\\
= & -\frac{1}{\pi}\left[\arg\langle\pi-\delta k\mid-\pi+\delta k\rangle+\theta(\pi-\delta k)-\theta(0)\right].\label{eq:oddCase}
\end{align}
\end{widetext}
The expression thereby becomes arbitrary. 
This even-odd subtlety vanishes when $m\rightarrow\infty$ and $\delta k\rightarrow0$, where the discrete Wilson polygon becomes a smooth loop. 
Note that Eq. \ref{eq:oddCase} remains gauge-invariant since the Wilson polygon is closed. Considering a gauge transformation $|k\rangle\rightarrow e^{i\chi(k)}|k\rangle$, we have $\theta(k)\rightarrow\theta(k)+\chi(k)-\chi(-k)$. Therefore
\begin{widetext}
\begin{align}
    & \arg\langle\pi-\delta k\mid-\pi+\delta k\rangle+\theta(\pi-\delta k)-\theta(0)\\
    \rightarrow & \arg\langle\pi-\delta k\mid-\pi+\delta k\rangle+[\cancel{\chi(-\pi+\delta k)}-\cancel{\chi(\pi-\delta k)}]+[\theta(\pi-\delta k)+\cancel{\chi(\pi-\delta k)}-\cancel{\chi(-\pi+\delta k)}]-\theta(0).
\end{align}
\end{widetext}

\section{Demonstration of Z2 index calculation for a full-band DFT model}

\begin{figure}
\begin{centering}
\includegraphics[width=1\columnwidth]{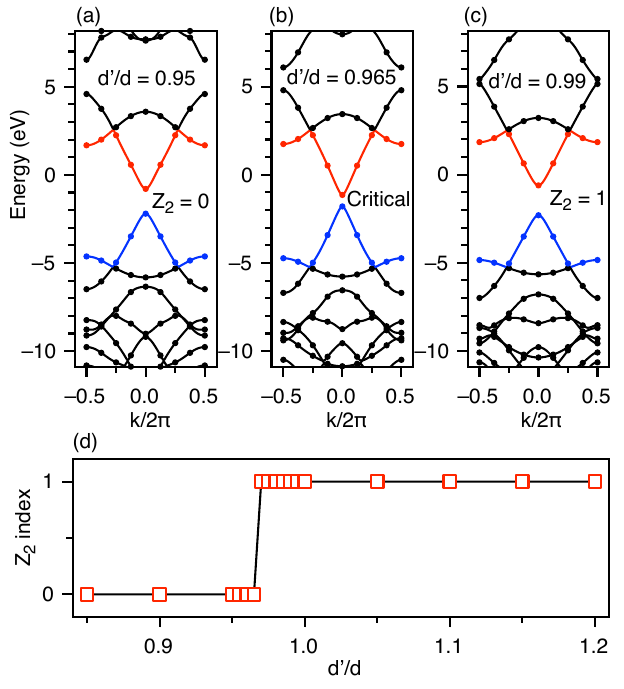}
\par\end{centering}
\caption{\textbf{DFT results and the $\mathcal{Z}_2$ transition.} (a-c) Three representative DFT band structures of [8]CPP performed in ORCA. The scatters denote the discrete molecular orbital energies, whereas the solid lines illustrate the smooth band structure due to the continuous $k$. The blue and red colors denote the two energy bands corresponding to topological band gap. (d) The transition of the $\mathcal{Z}_2$ index when changing the inter-unit C-C bond length $d'$ measured in the fixed C-C bond length within the benzene ring.   \label{fig:DFT_results}}
\end{figure}

Since the quantum manifold of a nanohoop stems from the discrete rotation symmetry, such manifold can be established using a converged density-functional theory (DFT) Hamiltonian if the AOs can be arranged respecting the symmetry. 
Also, the formalism introduced in our manuscript is assuming orthonormal basis. 
For the non-orthonormal AOs, one needs to orthonormalize the Hamiltonian using Cholesky decomposition. 
After these two steps, the Zak phase can be evaluated straightforwardly using Wilson polygon. 
To demonstrate this we have performed a series of DFT calculations for [8]CPP using the quantum chemistry package ORCA\cite{neese_software_2025} with the B3LYP functional\cite{becke_densityfunctional_1993,lee_development_1988} and the smallest Gaussian basis STO-3G\cite{hehre_selfconsistent_1972}.
We first restore the cyclic periodicity of the DFT results using Wigner D matrices. 
The orbitals on carbon are (in order) $1s$, $2s$, $2p_z$, $2p_x$, and $2p_y$. 
If we set the rotational symmetry axis of the [8]CPP to be the $z$-axis, we can use the following matrix block for each carbon atom.
    \begin{equation}
        r(\phi) = 
        \begin{bmatrix}
            1 & 0 & 0 & 0 & 0 \\
            0 & 1 & 0 & 0 & 0 \\
            0 & 0 & 1 & 0 & 0 \\
            0 & 0 & 0 & \cos(\phi) & \sin(\phi) \\
            0 & 0 & 0 & -\sin(\phi) & \cos(\phi)
        \end{bmatrix}.\label{eq:r_theta}
    \end{equation}
For each repeating unit (a single benzene ring), the appropriate matrix block is
    \begin{equation}
u(\phi) =
\begin{bmatrix}
    r(\phi) & 0        & \cdots & 0      & 0 \\
    0       & r(\phi)  & \cdots & 0      & 0 \\
    \vdots  & \vdots   & \ddots & \vdots & \vdots \\
    0       & 0        & \cdots & r(\phi)& 0 \\
    0       & 0        & \cdots & 0      & \mathbf I_4
\end{bmatrix},
\label{u_phi}
\end{equation}

i.e. we have one $r$ matrix for each carbon and the identity keeps the $s$ orbitals of the 4 hydrogens in place. Finally we have
\begin{equation}
    U =
    \begin{bmatrix}
        u(0)         & 0            & \cdots & 0 \\
        0            & u(\pi/4)     & \cdots & 0 \\
        \vdots       & \vdots       & \ddots & \vdots \\
        0            & 0            & \cdots & u(7\pi/4)
    \end{bmatrix}.
    \label{eq:rotating_U}
\end{equation}
For STO-3G this is a $272 \times 272$ matrix since $(6 \times 5 + 4) \times 8=272$. 
The cyclic periodic matrices can thus be restored as $\tilde{M}=UMU^\dagger$, where $M=F,S$, corresponding to the converged Fock and overlap matrices, respectively.  
We can construct $F(k)$ and $S(k)$ using a discrete Fourier transform, and then obtain the discrete `band structure' of the molecule, as illustrated in Figs. \ref{fig:DFT_results}(a-c). 
To trigger the $\mathcal{Z}_2$ transition, we performed a series of DFT calculation by changing the C-C bond length connecting adjacent repeating units. 
This can be parametrized by the ratio between the inter-unit bond length and the C-C bond length within each benzene ring: $\frac{d'}{d}$ where $d$ is fixed at $1.4\thinspace\textrm{\AA}$. 
The result is illustrated in Fig. \ref{fig:DFT_results}(d), where the HOMO-LUMO gap closing is synchronized with the sharp transition of the $\mathcal{Z}_2$ index. 

\section{Penetration depth of boundary states}

\begin{figure*}
\begin{centering}
\includegraphics[width=1\textwidth]{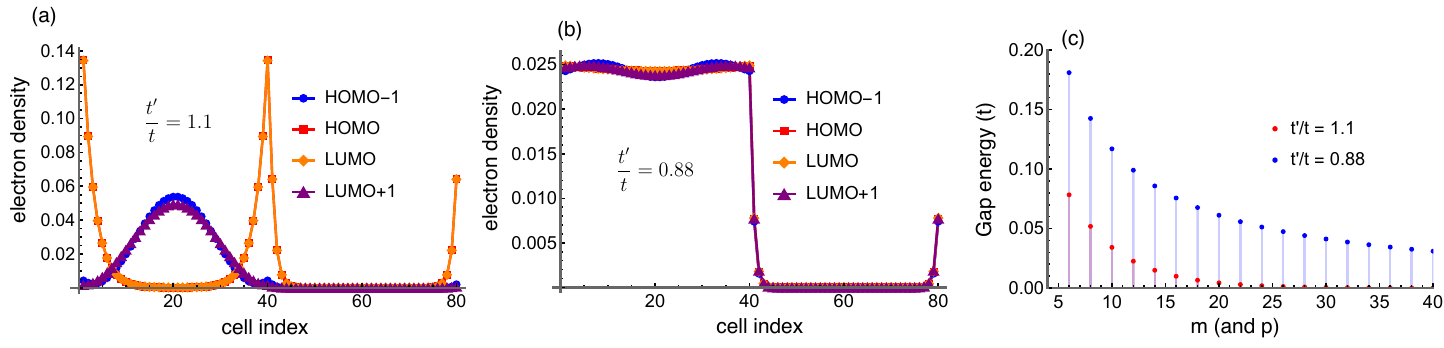}
\par\end{centering}
\caption{\textbf{The penetration depth of the boundary states and the scaling of the HOMO-LUMO gap} (a) The electron density per repeating unit for $\textrm{[m]-ITN-[p]-CPP}$ when $m=p=40$. (b) The electron density for the critical point near the phase transition $\frac{t'}{t}=0.88$, where the penetration depth diverges. (c) The scaling of the HOMO-LUMO gap as a function of $m$ for the two cases. \label{fig:penetrationDepth}}
\end{figure*}

The gap due to the interface coupling is mainly determined by the penetration depth of the boundary state. 
This depth is sensitively determined by the value of $\frac{t'}{t}$, as can be seen in Figs. \ref{fig:penetrationDepth}(a-b). 
Particularly, the case of $\frac{t'}{t}=1.1$ shown in Fig. \ref{fig:penetrationDepth}(a) corresponds to the nanohoop without any strain, whereas the one in Fig. \ref{fig:penetrationDepth}(b) is the stretched case near the topological transition. 
The later demonstrates the critical point of the phase transition where the electron density is simultaneously everywhere. 
The scaling of the HOMO-LUMO gap for these two cases are shown in Fig. \ref{fig:penetrationDepth}(c). 
We can see that the gap does not close when the molecule is stretched to the critical point (blue), regardless of the length of the two segments. 
However, the gap decreases exponentially as a function of $m$ (we assume $m=p$) without strain (red). 
This is consistent with the penetration depth of $\sim10$ on both sides, as shown in Fig. \ref{fig:penetrationDepth}(a). 
%


\end{document}